\documentclass[aps,pra,twocolumn,showpacs,prabib, superscriptaddress]{revtex4}
\usepackage{graphicx}
\usepackage{amsmath}
\usepackage{amssymb,wasysym}
\usepackage{amsfonts}
\usepackage{bm,verbatim}

\usepackage{color}

\newcommand{\bea}{\begin{eqnarray}}
\newcommand{\eea}{\end{eqnarray}}
\newcommand{\be}{\begin{equation}}
\newcommand{\ee}{\end{equation}}
\newcommand{\ds}{\displaystyle}
\newcommand{\rr}{\mathbf{r}}
\newcommand{\kk}{\mathbf{k}}

\newcommand{\RR}{\mathbf{R}}
\newcommand{\XX}{\mathbf{X}}
\newcommand{\JJ}{\mathbf{J}}
\newcommand{\SSS}{\mathbf{S}}
\newcommand{\cc}{\mathbf{c}}
\newcommand{\vn}{\mathbf{0}}
\newcommand{\Oom}{\mathbf{\Omega}}
\newcommand{\ra}{\rangle}
\newcommand{\la}{\langle}
\newcommand{\si}{\sigma}
\newcommand{\sip}{{\sigma'}}

\newcommand{\rhob}{\mbox{\boldmath$\rho$}}

\newcommand{\cA}{\mathcal{A}}
\newcommand{\Nr}{\mathcal{N}}

\newcounter{fnnumberbis}

\begin{document}

\title{General relations for quantum gases in two and three
dimensions.
\\ II. Bosons and mixtures}

\author{F\'elix Werner}
\affiliation{Department of Physics, University of Massachusetts,
Amherst, MA 01003, USA}
\affiliation{Laboratoire Kastler Brossel, \'Ecole Normale
Sup\'erieure, UPMC and CNRS, 24 rue Lhomond, 75231 Paris Cedex 05, France}

\author{Yvan Castin}
\affiliation{Laboratoire Kastler Brossel, \'Ecole Normale
Sup\'erieure, UPMC and CNRS, 24 rue Lhomond, 75231 Paris Cedex 05, France}

\begin{abstract}
We derive exact general relations between various observables for $N$ bosons with zero-range interactions,
in two or three dimensions, in an arbitrary external potential.
Some of our results are analogous to relations derived previously for two-component fermions,
and involve derivatives of the energy with respect to the two-body $s$-wave scattering
length $a$.
Moreover, in the three-dimensional case, where the Efimov effect takes place,
the interactions are characterized not only by $a$, but also by a three-body parameter $R_t$.
We then find additional relations which involve the derivative of the energy with respect $R_t$.
In short, this derivative gives the probability to find three particles close to each other.
Although it is evaluated for a totally loss-less model, it also gives the three-body
loss rate always present in experiments (due to three-body recombination to deeply bound diatomic molecules),
at least in the limit where the so-called inelasticity parameter $\eta$ is small enough.
As an application, we obtain, within the zero-range model and to first order in $\eta$,
an analytic expression
for the three-body loss rate constant for a non-degenerate Bose gas at thermal equilibrium with infinite scattering length.
 We also discuss the generalization to arbitrary mixtures of  bosons and/or fermions.
  \end{abstract}

%
\date{\today}

\maketitle

\section{Introduction}

Ultracold atomic gases with resonant interactions, that is having a $s$-wave scattering length much larger in absolute
value than the interaction range, can now be studied experimentally thanks to the broad magnetic Feshbach resonances,
not only with two-component fermions \cite{Varenna,ZwergerLivre} but also with bosons \cite{Cornell_kFa=1,Salomon_Bose_strong,CornellCBose,
GrimmRevue,KhaykovichTrimeres} 
or mixtures \cite{JochimEfimovRF,NaidonTrimeres}.
In this resonant regime, one can neglect the range of the interaction, which is equivalent to replacing
the interaction with contact conditions on the $N$-body wavefunction: In 3D, this constitutes the so-called
zero-range model
\cite{Efimov,AlbeverioLivre,YvanHouchesBEC,PetrovPRL,PetrovJPhysB,RevueBraaten,WernerThese},
that can also be defined in 2D (see e.g.\ \cite{Busch,PetrovShlyapCollisions2D,MaximLudo2D,LudoScatteringLowD}),
and of course in 1D \cite{LiebLiniger,GaudinLivre}. In each dimension, these models include a length, the so-called
$d$-dimensional scattering length $a$. In three dimensions, when the Efimov effect occurs~\cite{Efimov}, an additional
length has to be introduced, the so-called three-body parameter \cite{Danilov}.

For the zero-range models, it was gradually realized that several observables, such as the short distance behavior of the pair
distribution function $g^{(2)}(\rr)$ or the tail of the momentum distribution $n(\kk)$, 
can be related to derivatives of the energy with respect to the $d$-dimensional scattering length $a$.
In 1D, the value of $g^{(2)}(0)$ was directly related to such a derivative
by the Hellmann-Feynman theorem \cite{LiebLiniger}; the coefficient of the leading $1/k^4$ term
in $n(k)$ at large $k$ was then related to the singular behavior of the wavefunction for two close particles,
and ultimately to $g^{(2)}(0)$, by general properties of the Fourier transform \cite{Olshanii_nk}.
In 3D, for spin-$1/2$ fermions (where the Efimov effect does not occur), an extension of the 1D relations
was obtained by a variety of techniques \cite{TanEnergetics,TanLargeMomentum,Braaten,BraatenLong,
WernerTarruellCastin,ZhangLeggettUniv}, including the original 1D techniques. Generalizations were then obtained
for 2D systems, for fermions or bosons \cite{TanSimple,CombescotC,Moelmer,CompanionFermions}.

This is the second of a series of two articles on such general relations.
The first one covered two-component fermions (Ref.~\cite{CompanionFermions}, hereafter referred to as Article~I).
Here, we consider single-component bosons, as well as mixtures.
In the 3D case, remarkably, the Efimov effect leads to modifications or even breakdown of some relations, 
and to the appearance of additional relations involving the derivative of the energy with respect to
the three-body parameter $R_t$. Several of the results presented here were already contained in~\cite{50pages}
and rederived in ~\cite{BraatenBosons} with a different technique, that allowed the authors of~\cite{BraatenBosons}
to obtain still other Efimovian relations for $N$ bosons \footnote{The ``three-body contact'' parameter $C_3$ of~\cite{BraatenBosons} 
is equal to $(\partial_{\ln R_t} E)_a m/(2\hbar^2)$ in our notations.}.

The article is organized as follows.
Section~\ref{sec:intro} introduces the zero-range model and associated notations for the single-component bosons.
Section~\ref{sec:analog} presents relations which are analogous to the fermionic ones.
Additional relations resulting from the Efimov effect are derived in Section~\ref{sec:new_rel}.
As an application, the three-body loss rate of a non-degenerate Bose gas for an infinite scattering length
is calculated in Section~\ref{sec:tblr}.
Finally the case of an arbitrary mixture is addressed in Section \ref{sec:melange}.
We conclude in Section \ref{sec:conclusion}.
Note that, for convenience, the main relations are displayed in Tables~I, II, III.

\section{Model and notations}
\label{sec:intro}

In 3D, the zero-range model imposes the Wigner-Bethe-Peierls contact condition on the $N$-body wavefunction:
For any pair of particles $i,j$, when one takes the limit of a vanishing distance $r_{ij}\equiv|\rr_i-\rr_j|$ with a
fixed value of the center of mass $\cc_{ij}=(\rr_i+\rr_j)/2$ different from the positions $\rr_k$ of the 
other $N-2$ particles, the wavefunction has to behave as
\be
\psi(\rr_1,\ldots,\rr_N)= \left(\frac{1}{r_{ij}}-\frac{1}{a}\right)A_{ij}(\cc_{ij},(\rr_k)_{k\neq i,j})+O(r_{ij})
\label{eq:wbp3d}
\ee
where $a$ is the 3D scattering length. The {\sl a priori} unknown functions $A_{ij}$ are determined
from the fact that $\psi$ solves the free Schr\"odinger's equation over the domain where the positions
of the particles are two by two distinct: $E\psi = H\psi$ with
\be
H=\sum_{i=1}^{N} \left[-\frac{\hbar^2}{2m} \Delta_{\rr_i} + U(\rr_i)\right]
\label{eq:hamil}
\ee
and $U$ is the external potential.
Also $\psi$ is normalized to unity.

If there are three bosons or more, the Efimov effect occurs~\cite{Efimov}, and the zero-range model has to be supplemented
by a three-body contact condition that involves a positive length, the three-body parameter $R_t$:
In the limit where {\sl three} particles approach each other (that one can take to be particles  1, 2 and 3 due
to the bosonic symmetry),
there exists a function $B$, hereafter called three-body regular part, such that
\be
\psi(\rr_1,\ldots,\rr_N)\underset{R\to0}{\sim}\Phi(\rr_1,\rr_2,\rr_3)\, B(\cc_{123},\rr_4,\ldots,\rr_N)
\label{eq:danilov}
\ee
where 
$\cc_{123}=(\rr_1+\rr_2+\rr_3)/3$ is the center of mass of particles $1$,$2$ and $3$,
$\Phi$ is the zero-energy three-body scattering state
\be
\Phi(\rr_1,\rr_2,\rr_3) = \frac{1}{R^2}\sin\left[|s_0|\ln\frac{R}{R_t}\right] \phi_{s_0}(\Oom),
\label{eq:Phi_ZR}
\ee
and where $R, \Oom$ are the hyperradius and the hyperangles associated with particles $1$, $2$ and $3$.
We take the limit $R\to0$ in~(\ref{eq:danilov}) for fixed $\Oom$ and $\cc_{123}$ (in analogy with the two-body contact condition).

We recall the definition of $R$ and $\Oom$: From the Jacobi coordinates $\rr=\rr_2-\rr_1$ and 
$\rhob=(2\rr_3-\rr_1-\rr_2)/\sqrt{3}$, one forms the six-component vector 
$\RR=(\rr,\rhob)/\sqrt{2}$; then, the hyperradius $R=\sqrt{(r^2+\rho^2)/2}$ is the norm
of $\RR$, and $\Oom=\RR/R$ is its direction that can be parametrized by five 
hyperangles, so that $d^6R=R^5 dR d^5\Omega$.
In Eq.~(\ref{eq:Phi_ZR}), $s_0=i\cdot 1.00623782510\ldots$ is Efimov's transcendental number, 
it is the imaginary solution (with positive imaginary part)
of $s \cos(s \pi/2) = (8/\sqrt{3}) \sin( s \pi / 6)$;
$\phi_{s_0}(\Oom)$ is the hyperangular part of the Efimov trimers wavefunctions~\cite{Efimov}, which, in the present case (single-component bosons), is given by 
$\phi_{s_0}(\Oom)\equiv \Nr\,(1+Q)\sinh\left[|s_0|\left(\frac{\pi}{2}-\alpha\right)\right]
/\sin(2\alpha)$ where $Q=P_{13}+P_{23}$ and $P_{ij}$ exchanges particles $i$ and $j$,
and where $\alpha\equiv {\rm arctan}(r/\rho)$.
Here we introduced, for later convenience, a normalization factor
such that $\int d^5\Omega\, |\phi_{s_0}(\Oom)|^2 = 1$. Using
$\int d^5\Omega = \int_0^{\pi/2}d\alpha\sin^2\alpha\cos^2\alpha
\int d^2\hat{r}\int d^2\hat{\rho}$, where $d^2\hat{r}$ and $d^2\hat{\rho}$ are the 
differential solid angles in 3D,  we obtain ~\cite{Efimov93,CastinWernerTrimere}
\begin{multline}
\Nr^{-2} = \frac{6\pi^2}{|s_0|} \sinh(|s_0|\pi/2) \Big[\cosh(|s_0|\pi/2) 
\\
+ |s_0| \frac{\pi}{2} \sinh (|s_0|\pi/2) -\frac{4\pi}{3\sqrt{3}}\cosh (|s_0|\pi/6)\Big] .
\end{multline}

For $N=3$ particles, it is well established that this model Hamiltonian is self-adjoint and that it is the zero-range limit of finite-range models, 
see e.g.~\cite{WernerThese} and references therein.
The fact that the zero-range (i.e. low-energy) regime can be described using the scattering length and a three-body parameter only is known as universality~\cite{RevueBraaten}.
For $N=4$, an accurate numerical study \cite{Deltuva} has shown, as was suggested by earlier ones 
\cite{Stecher4corps,Hammer4corps1,Hammer4corps2} and as supported by experimental evidence \cite{Grimm4body}, that there is no need to introduce a 
four-body parameter in the zero-range limit,
implying that the here considered zero-range model Hamiltonian is self-adjoint for $N=4$.
Physically, this is related to the fact that the introduction of $R_t$, 
imposed by the three-body Efimov effect,
necessarily breaks the separability of the 4-body problem at infinite scattering length ; this precludes the 
simplest scenario imposing the introduction of a four-body parameter, namely 
a four-body Efimov effect such as the one found for $3+1$ fermions in \cite{CMP}.
Here we consider an arbitrary value of $N$ such that the model Hamiltonian is self-adjoint.

In 2D, the zero-range model is a direct generalization of the 3D one, since one simply replaces the 3D zero-energy
two-body scattering wavefunction $r_{ij}^{-1}-a^{-1}$ by the 2D one $\ln(r_{ij}/a)$, where $a$ is now the 2D scattering
length. Accordingly, for any pair of particles $i$ and $j$, in the limit $r_{ij}\equiv |\rr_i-\rr_j|\to 0$ with $\cc_{ij}=(\rr_i+\rr_j)/2$ fixed,
the $N$-body wavefunction satisfies in 2D:
\be
\psi(\rr_1,\ldots,\rr_N)= \ln(r_{ij}/a)A_{ij}(\cc_{ij},(\rr_k)_{k\neq i,j})+O(r_{ij}).
\label{eq:wbp2d}
\ee
There is no Efimov effect in 2D so that no additional parameter is required 
\cite{BruchTjon3bosons2D,Fedorov3bosons2D,Leyronas4corps}.
The Hamiltonian is the corresponding 2D version of (\ref{eq:hamil}).

\section{Relations which are analogous to the fermionic case}
\label{sec:analog}


\begin{table*}
\begin{tabular}{|cc|cc|}
\hline  
  Three dimensions & & Two dimensions & \\
\hline
 \multicolumn{4}{|c|}{\vspace{-4mm}} 
\\
 \multicolumn{3}{|c}{${C}\equiv {\displaystyle \lim_{k\to +\infty}} 
k^4 n(\kk)$}
& (1) \vspace{-4mm}
\\
  \multicolumn{4}{|c|}{} \\
\hline 
\vspace{-4mm}
& & & \\
\vspace{-4mm}
$\ds {C} =
32\, \pi^2\ (A,A)
$
&
(2a)
&
$\ds {C} = 
8\, \pi^2\,(A,A)
$
&
(2b)
\\
& & & \\
\hline
& & &
\vspace{-4mm}
\\
$\ds
\int d^3c \,
 g^{(2)} \left(\mathbf{c}+\frac{\mathbf{r}}{2},
\mathbf{c}-\frac{\mathbf{r}}{2}\right) 
\underset{r\to0}{\sim}
\frac{{C}}{(4\pi)^2} 
\frac{1}{r^2}
$
&(3a)  \vspace{-4mm}
&
$\ds
\int d^2c \,
 g^{(2)} \left(\mathbf{c}+\frac{\mathbf{r}}{2},
\mathbf{c}-\frac{\mathbf{r}}{2}\right)
\underset{r\to0}{\sim}
\frac{{C}}{(2\pi)^2} 
\ln^2 r$
&
(3b)
\\
& & & 
\\
 \hline & & &
\vspace{-4mm}
\\
$\displaystyle\left(\frac{\partial E}{\partial(-1/a)}\right)_{\!R_t} = \frac{\hbar^2 {C}}{8\pi m} $ &
 (4a) &
$\ds\frac{dE}{d(\ln a)} = \frac{\hbar^2 {C}}{4\pi m} $ 
& (4b)  \vspace{-4mm}
 \\
&& & \\
\hline & & & \vspace{-4mm} \\
$\ds E - E_{\rm trap}  \stackrel{\mathrm{if}\, \exists\, \mathrm{lim}}{=} \frac{\hbar^2 {C}}{8\pi m a}  $ 
&  &
$\ds E - E_{\rm trap}  = \lim_{\Lambda\to\infty}\left[-\frac{\hbar^2 {C}}{4\pi m} \ln \left(\frac{a \Lambda e^\gamma}{2}\right) \right.
$
&
\vspace{-3mm}
\\
& & & \\ 
$\ds + \lim_{\Lambda\to\infty}\int_{k<\Lambda} \frac{d^3\!k}{(2\pi)^3}  \frac{\hbar^2 k^2}{2m} 
\left[n(\kk) - \frac{{C}}{k^4}\right]$
&(5a)
&
$\ds  + \left. \int_{k<\Lambda} \frac{d^2\!k}{(2\pi)^2}  \frac{\hbar^2 k^2}{2m} 
n(\kk) \right]$
& (5b)\vspace{-4mm}
 \\
&& & \\
\hline 
& & & \vspace{-8mm} \\
 \\$\ds\frac{1}{2} \left(\frac{\partial^2E_n}{\partial(-1/a)^2}\right)_{\!R_t}
= \left(\frac{4\pi\hbar^2}{m}\right)^2 \sum_{n',E_{n'}\neq E_n} 
\frac{|(A^{(n')},A^{(n)})|^2}{E_n-E_{n'}}$
&(6a)  &
$\ds\frac{1}{2} \frac{d^2E_n}{d(\ln a)^2}
= \left(\frac{2\pi\hbar^2}{m}\right)^2 \sum_{n',E_{n'}\neq E_n} 
\frac{|(A^{(n')},A^{(n)})|^2}{E_n-E_{n'}} $
&
 (6b) \vspace{-4mm}
 \\&&
& \\
 \hline
\end{tabular}
\caption{For single-component bosons, relations which are analogous to the fermionic case.
In three dimensions, the derivatives are taken for a fixed three-body parameter $R_t$.
As discussed in the text, in three dimensions, the relation between energy and momentum distribution is valid if the  large cut-off limit $\Lambda\to +\infty$ exists, which is not the case for Efimovian states
(i.e.\ eigenstates whose energy depends on $R_t$). The notation $(A,A)$ is defined in Eq.~(\ref{eq:AA}). $\gamma=0.577215\ldots$ is Euler's constant.
\label{tab:bosons1}}
\end{table*}

A first set of relations is given in Table~\ref{tab:bosons1}. These relations and derivations are largely analogous to the fermionic case (which was treated in Article~I). 
An obvious difference with the fermionic case is that there are no more spin indices in the pair distribution function $g^{(2)}$ and in the momentum distribution $n(\kk)$. Accordingly
we now have $g^{(2)}(\rr,\rr')=\langle
\hat{\psi}^\dagger(\rr)
\hat{\psi}^\dagger(\rr')
\hat{\psi}(\rr')
\hat{\psi}(\rr)
\rangle
=
\int d^d r_1\ldots d^d r_N 
\left|
\psi(\rr_1,\ldots,\rr_N)
\right|^2
\sum_{i\neq j}
\delta\left(\rr-\rr_i\right)
\delta\left(\rr'-\rr_j\right)
$,
where $\hat{\psi}$ is the bosonic field operator,
and the momentum distribution is normalized as
$
\int  n(\kk) d^d k/(2\pi)^d= N
$.
Apart from numerical prefactors, there are two more important differences which appear in the 3D case due to the Efimov effect.

The first important difference is that the derivatives with respect to $1/a$ in [Tab.~\ref{tab:bosons1}, Eqs.~(4a,6a)] have to be taken for a fixed three-body parameter $R_t$.
This comes from the relation 
\be
\left( \frac{\partial E}{\partial(-1/a)} \right)_{\!R_t} = \frac{4\pi\hbar^2}{m} (A, A)
\label{eq:AA_3D}
\ee
with the notation (given for generality in dimension $d$):
\be
(A,A)\equiv \sum_{i<j} \int (\prod_{k\neq i,j}d^d r_k)\int d^dc_{ij} |A_{ij}(\cc_{ij},(\rr_k)_{k\neq i,j})|^2.
\label{eq:AA}
\ee
Eq.~(\ref{eq:AA_3D}) was already obtained in~\cite{WernerThese} in the case $N=3$.
A simple way to derive it for any $N$ is to use a cubic lattice model, of lattice spacing $b$, with purely on-site interactions 
characterized by a coupling constant $g_0$ [see the Hamiltonian in Eq.~(\ref{eq:Hlatt}) below, with $h_0=0$], 
adjusted to reproduce the correct scattering length \cite{MoraCastin}:
\be
\frac{1}{g_0}= \frac{m}{4\pi\hbar^2 a} -\int_{D} \frac{d^3k}{(2\pi)^3} \frac{m}{\hbar^2 k^2}
\label{eq:val_g0}
\ee
where the wavevector $\kk$ of a single-particle plane wave on the lattice
is restricted as usual to the first Brillouin zone $D=(-\frac{\pi}{b},\frac{\pi}{b})^3$.
One then follows the same reasoning as in (Article~I, Section~V, Subsections~C-D-E). The key point here is that, 
in the limit of $b\ll |a|$, the three-body parameter corresponding to the lattice model is equal to a numerical constant times $b$~\footnote{The value of this constant is irrelevant for what follows. It could be calculated e.g.\ by equating the energies of the weakly bound Efimov trimers of the lattice model with the ones of the zero-range model. This was done e.g.\ in~\cite{Werner3corpsPRL,WernerThese}, not for the lattice model, but for a Gaussian separable potential model.}. Thus, varying the coupling constant $g_0$ while keeping $b$ fixed
is equivalent to varying $a$ while keeping $R_t$ fixed, so that
\be
\frac{dE}{dg_0}=\left(\frac{dE}{d(-1/a)}\right)_{\!R_t}\ \frac{d(-1/a)}{dg_0}.
\label{eq:dEdg0_Rt}
\ee
The left-hand side of (\ref{eq:dEdg0_Rt}) is given by the Hellmann-Feynman theorem: 
\begin{multline}
\frac{dE}{dg_0} = \frac{1}{2} \sum_{\rr} b^3 \langle (\hat{\psi}^\dagger \hat{\psi}^\dagger \hat{\psi}\hat{\psi})(\rr)\rangle \\
=\frac{N(N-1)}{2} \sum_{\rr,\rr_3,\ldots,\rr_N} b^{3(N-1)} |\psi(\rr,\rr,\rr_3,\ldots,\rr_N)|^2
\end{multline}
where $\psi$ is the eigenstate wavefunction on the lattice. In the zero-range limit $b\ll |a|$, $\psi$ has to match
the contact condition (\ref{eq:wbp3d}): Its two-body regular part $A_{12}$, defined as
\be
\psi(\rr,\rr,\rr_3,\ldots,\rr_N)\equiv \phi(\mathbf{0}) A_{12}(\rr,\rr_3,\ldots,\rr_N),
\ee
with the correctly normalized zero-energy two-body lattice scattering wavefunction $\phi(\rr)$ [$\phi(\rr)=r^{-1}-a^{-1}+o(1)$ at $r\gg b$],
has to converge to the zero-range model regular part.
Similarly, in the right-hand side of (\ref{eq:dEdg0_Rt}), the lattice model's $(dE/d(-1/a))_{R_t}$ tends to the zero-range model's one 
if one takes the zero-range limit while keeping $R_t$ fixed~\footnote{The zero-range limit for a fixed $R_t$ can be taken by repeatedly dividing $b$ by the discrete scaling factor $\exp(\pi/|s_0|)$
 and by adjusting $g_0$ so that $a$ remains fixed. In this limit the ground state energy tends to $-\infty$ as follows from the Thomas effect, but 
the restriction of the spectrum to any fixed energy window converges (see e.g.~\cite{WernerThese}).}.
\setcounter{fnnumberbis}{\thefootnote}
Finally, the last factor of (\ref{eq:dEdg0_Rt}) can be evaluated from (\ref{eq:val_g0}).
Using the relation $\phi(\mathbf{0}) =-4\pi\hbar^2/(m g_0)$ established in \cite{CompanionFermions}, we obtain
Eq.~(\ref{eq:AA_3D}).
The same lattice model reasoning explains why the second-order derivative in [Tab.~\ref{tab:bosons1}, Eq.~(6a)]
also has to be taken for a fixed $R_t$.

The second important difference with respect to the fermionic case is that the relation [Tab.~\ref{tab:bosons1}, Eq.~(5a)] breaks down in general, and only holds for special states for which the infinite-cutoff limit $\Lambda\to\infty$ exists (such as the universal states for 3 trapped bosons of~\cite{Pethick3corps,Werner3corpsPRL}).
This was overlooked in~\cite{CombescotC}, and was shown for an Efimov trimer in~\cite{CastinWernerTrimere}.
The correct relation valid for any $N$-body state in presence of the Efimov effect was obtained in~\cite{BraatenBosons}.

\bigskip
{\bf Note added by Y. Castin after publication:} 
In the fermionic case, it was shown in 2D and in 3D in \cite{CompanionFermions} that the first correction
to any eigenenergy $E$ due to a non-zero range of the interaction is of the form $B r_e^2$ in 2D and $B r_e$ in 3D,
where $r_e$ is the effective range of the interaction and the coefficient $B$ is universal (it depends on the considered eigenstate but
not on the interaction potential). For 2D bosons, where there is no Efimov effect, we expect the same property to hold, and we illustrate
it here briefly by a calculation of the coefficient $B$ for the ground state of the weakly interacting Bose gas in the Bogoliubov 
approximation. 
 
We reuse the lattice model and the Bogoliubov results of \cite{MoraCastin}, 
or more conveniently of C. Mora, Y. Castin [Phys. Rev. Lett. {\bf 102}, 180404 (2009)]. The zero-temperature grand potential $\Omega$
of the spatially homogeneous weakly interacting Bose gas of area $L^2$ is given in the thermodynamic limit by
\[
\Omega_{\rm Bog} L^{-2} = -\frac{\mu^2}{2g_0}-\int_{D} \frac{d^2k}{(2\pi)^2} \epsilon_k V_k^2
\]
Here $b$ is the square lattice spacing and $D=[-\pi/b,\pi/b[^2$ the corresponding first Brillouin zone, $g_0$ is the bare coupling constant 
giving the on-site interaction $g_0 b^{-2} \delta_{\rr_i,\rr_j}$ between bosons $i$ and $j$, $\mu$ is the chemical potential, 
$\epsilon_k=[E_k(E_k+2\mu)]^{1/2}$ is the Bogoliubov excitation spectrum at wavevector $\kk$ (with $E_k=\hbar^2 k^2/2m$) and the Bogoliubov
mode amplitudes $U_k$ and $V_k$ are given by $U_k+V_k=(U_k-V_k)^{-1}=[E_k/(E_k+2\mu)]^{1/4}$. The trick is then to introduce the two-body
$T$ matrix of the lattice model, given by Eq.~(164) in \cite{MoraCastin}, that depends only on the energy, not on the incoming
and outgoing relative wavevectors:
\bea
\langle \kk |T(E+i0^+)|\kk'\rangle &=& t_0(E+i0^+) \ \forall E\in \mathbb{R} \notag\\
\frac{1}{t_0(z)} &=&  \frac{1}{g_0} - \int_{D} \frac{d^2k}{(2\pi)^2} \frac{1}{z-2E_k} 
\forall z\in \mathbb{C}\setminus{\mathbb{R}^+} \notag
\eea
where the index $0$ in $t_0$ makes it apparent that this is pure $s$-wave scattering.  
Then one eliminates $1/g_0$ in terms of the $T$ matrix at energy $-2\mu$:
\[
\Omega_{\rm Bog} L^{-2} = -\frac{\mu^2}{2 t_0(-2\mu)} - \int_{D} \frac{d^2k}{(2\pi)^2} \left[\epsilon_k V_k^2 -
\frac{\mu^2}{4(E_k+\mu)}\right]
\label{eq:astuce}
\]
It remains to take the zero lattice spacing limit $b/\xi\to 0$ where $\xi$ is the healing length such that $\hbar^2/m\xi^2=\mu$.
In the expression above, the integral deviates from its $b\to 0$ limit (obtained by replacing the first Brillouin zone by $\mathbb{R}^2$)
by $O(m\mu^2 (b/\xi)^4/\hbar^2)$, 
since its integrand is $O(\mu^4/E_k^3)$ when $k\to +\infty$; this deviation is $O(r_e^4)$ and negligible. 
On the contrary, $t_0(-2\mu)$ varies to second order in $b$. 
By definition of the effective range $r_e$, see Eqs.~(96,97) in \cite{CompanionFermions}, one has for $E\to 0^+$:
\[
t_0(E+i0^+)=\frac{-4\pi \hbar^2/m}{-i\pi+\ln\frac{m E e^{2\gamma}a^2}{4\hbar^2}+ \frac{m E r_e^2}{\hbar^2} + O(\frac{m^2E^2 b^4}{\hbar^4})}
\]
where $\gamma=0.577\, 215\ldots$ is Euler's constant, $a$ the 2D scattering length. We find for the lattice model $r_e^2=
\left(\frac{b}{\pi}\right)^2 \left(\frac{1}{\pi}+\frac{1}{2}\right)$. In conclusion,
to first order in $r_e^2$, the density of grand potential for the weakly interacting 2D Bose gas is predicted by the grand canonical
Bogoliubov theory to be given by the universal formula
\[
\Omega_{\rm Bog} L^{-2} = \frac{m\mu^2}{8\pi\hbar^2} \ln \frac{m\mu a^2 e^{2\gamma+1/2}}{4\hbar^2} 
-\frac{m^2\mu^3}{4\pi\hbar^4} r_e^2
\]
We recall that the universality of the $\propto r_e^2$ energy correction predicted in \cite{CompanionFermions} 
holds in the zero-range limit $b\to 0$. If one is not strictly in this limit, $\Omega L^{-2}$ was expected in \cite{CompanionFermions}
to exhibit a second corrective term, proportional to the square of the {\sl off-shell} effective range $\rho_e$,
see Eq.~(106) in \cite{CompanionFermions}, whereas $r_e$ is here the {\sl on-shell} effective range.
In general, both $r_e^2$ and $\rho_e^2$ are of order $b^2$, where $b$ is the true interaction range, 
so the $\propto \rho_e^2$ and the $\propto r_e^2$ corrective terms to $\Omega$ seem to be of the same order of magnitude.
This paradox was clarified in \cite{CompanionFermions}, where it was shown for 
short-range interaction potentials $V(r)$ that in the zero-range resonant-scattering
limit $b\to 0$ at fixed $a$, $\rho_e^2/r_e^2 = O(1/\ln(a/b))\to 0$ so that the $\propto\rho_e^2$ correction becomes negligible. This resonant-scattering
limit is however not desirable with bosons, if one wishes to avoid the formation of bound states. It is in particular
not achieved in the popular hard disk model. In the lattice model of \cite{MoraCastin} that we used here,
one has exactly $\rho_e=0$ and the $\propto\rho_e^2$ correction does not show up and cannot be evaluated.  {\bf End of the added note}.

\section{Additional relations coming from the Efimov effect}
\label{sec:new_rel}

\begin{table*}
\begin{tabular}{|cc|}
\hline
\vspace{-4mm}
&
\\
 $\ds \left(\frac{\partial E}{\partial (\ln R_t)}\right)_{\!\!a}=\frac{\hbar^2}{m}\,\frac{\sqrt{3} \,|s_0|^2}{4} N(N-1)(N-2)
\int d^3c_{123}\, d^3r_4\ldots d^3r_N\,|B(\cc_{123},\rr_4,\ldots,\rr_N)|^2$
& (1)
\\
\vspace{-4mm}
&
\\
\hline
\vspace{-4mm}
&
\\
$\ds \int d^3c_{123}\,g^{(3)}(\rr_1,\rr_2,\rr_3) \underset{R\to0}{\sim}
|\Phi(\rr_1,\rr_2,\rr_3)|^2
\,\left(\frac{\partial E}{\partial (\ln R_t)}\right)_{\!\!a}\ 
\frac{4}{\sqrt{3}\,|s_0|^2}
\,\frac{m}{\hbar^2}$
& (2)
\\
\vspace{-4mm}
&
\\
\hline
\vspace{-4mm}
&
\\
$\ds \hbar\Gamma \underset{\eta\to 0}{\sim} \left(\frac{\partial E}{\partial (\ln R_t)}\right)_{\!\!a}\  \frac{2\eta}{|s_0|}$ & (3)
\vspace{-4mm}
 \\&
 \\
 \hline
\end{tabular}
\caption{For single-component bosons in 3D, additional relations coming from the Efimov effect.
 $B$ is the three-body regular part of the $N$-body wavefunction; $g^{(3)}$ is the triplet distribution function;
 $\Gamma$ is the decay rate due to three-body losses and $\eta$ is the corresponding inelasticity parameter
 (see text). The integral in (2) is taken for fixed relative coordinates.
\label{tab:bosons2}}
\end{table*}

In addition to modifying relations which already existed for fermions,
the Efimov effect gives rise to additional relations, involving the
derivative of the energy with respect to the logarithm of the three-body parameter.
 These relations are displayed in Table~\ref{tab:bosons2}.

\subsection{Derivative of the energy with respect to the three-body parameter}

Our first additional relation~[Tab.~\ref{tab:bosons2}, Eq.~(1)] expresses 
the derivative of the energy with respect to the three-body parameter $R_t$ in terms
of the three-body regular part defined in Eq.~(\ref{eq:danilov}).
This is similar to the relation 
(\ref{eq:AA_3D})
between the derivative with respect to the scattering length and the (two-body) regular part
\footnote{We note that it was already speculated in \cite{Braaten} that, in presence of the Efimov effect, ``a three-body analog of the contact'' may ``play an important role''.}. 
We will first derive this relation using the zero-range model in the case $N=3$, and then using a lattice model for any $N$.

\subsubsection{Derivation using the zero-range model for three particles}
We consider two wavefunctions $\psi_1$, $\psi_2$, satisfying the two-body 
boundary condition (\ref{eq:wbp3d})
with the same scattering length $a$, and the three-body boundary condition 
(\ref{eq:danilov},\ref{eq:Phi_ZR}) 
with different three-body parameters $R_{t 1}$, $R_{t 2}$. The corresponding three-body regular parts are denoted by $B_1$, $B_2$.
We show in the Appendix~\ref{app:3b} that
\begin{multline}
\la \psi_1, H \psi_2\ra-\la H\psi_1,\psi_2\ra=\frac{\hbar^2}{m}\frac{3\sqrt{3}|s_0|}{2}\sin\left[|s_0|\ln\frac{R_{t 2}}{R_{t 1}}\right]\,
\\ \times \int d^3c_{123}\,B^*_1(\cc_{123})B_2(\cc_{123}),
\label{eq:lemme_dEdRt}
\end{multline}
which yields [Tab.~II, Eq.~(1)] by choosing $\psi_i$ as an eigenstate of energy $E_i$ and taking the limit $R_{t 2}\to R_{t 1}$
\footnote{We note that $\psi_1$ and $\psi_2$ do not satisfy the lemma [Article~I, Eq.~(33)] because they are too singular for $R\to0$.
If this lemma was applying, the right-hand side of (\ref{eq:lemme_dEdRt}) would be zero and the two-body contact condition
(\ref{eq:wbp3d}) would define a self-adjoint Hamiltonian without need of the extra, three-body contact condition (\ref{eq:danilov}),
which is not the case.}.

\subsubsection{Derivation using a lattice model}

We now derive [Tab.~II, Eq.~(1)] for all $N$ using as in Sec.~\ref{sec:analog} a cubic lattice model, 
except that the Hamiltonian now contains a three-body interaction term (of coupling constant $h_0$)
allowing one to adjust the three-body parameter $R_t$ without changing the lattice spacing:
\begin{multline}
H_{\rm latt}=\int_D \frac{d^3k}{(2\pi)^3}\, \frac{\hbar^2 k^2}{2m} \hat{c}^\dagger(\kk)\hat{c}(\kk) + \sum_{\rr} b^3 U(\rr) (\hat{\psi}^\dagger 
\hat{\psi})(\rr) 
\\+\frac{g_0}{2} \sum_\rr b^3 (\hat{\psi}^\dagger\hat{\psi}^\dagger\hat{\psi}\hat{\psi})(\rr)
+h_0 \sum_\rr b^3 (\hat{\psi}^\dagger\hat{\psi}^\dagger\hat{\psi}^\dagger\hat{\psi}\hat{\psi}\hat{\psi})(\rr).
\label{eq:Hlatt}
\end{multline}
Here the bosonic field operator obeys discrete commutation relations 
$[\hat{\psi}(\rr),\hat{\psi}^\dagger(\rr')]=\delta_{\rr\rr'}/b^3$ and the plane wave annihilation operators
obey as usual $[\hat{c}_\kk,\hat{c}^\dagger_{\kk'}]=(2\pi)^3\delta(\kk-\kk')$ provided that $\kk$ and $\kk'$ are restricted to
the first Brillouin zone $D$.

We then define the zero-energy three-body scattering state $\phi_0(\rr_1,\rr_2,\rr_3)$
as the solution of $H_{\rm latt}|\phi_0\ra=0$ for $a=\infty$, with the boundary condition
\be
\phi_0(\rr_1,\rr_2,\rr_3)\sim \Phi(\rr_1,\rr_2,\rr_3)
\ee
in the limit where all interparticle distances tend to infinity.
Here $\Phi$ is the zero-range model's zero-energy scattering state, given in Eq.~(\ref{eq:Phi_ZR}).
This defines the three-body parameter $R_t(b,h_0)$ for the lattice model (since $\Phi$ depends on $R_t$).
The Hellmann-Feynman theorem writes:
\begin{multline}
\frac{\partial E}{\partial h_0} = \sum_{\rr} b^3\,\la (\psi^\dagger\psi^\dagger\psi^\dagger\psi\psi\psi)(\rr)\ra
\\ =
N(N-1)(N-2)\sum_{\rr,\rr_4,\ldots,\rr_N}b^{3(N-2)}|\psi(\rr,\rr,\rr,\rr_4,\ldots,\rr_N)|^2.
\end{multline}
For the lattice model we define the three-body regular part $B$ through:
\be
\psi(\rr,\rr,\rr,\rr_4,\ldots,\rr_N)=\phi_0(\vn,\vn,\vn)\,B(\rr,\rr_4,\ldots,\rr_N);
\ee
in the zero-range limit, we expect that this lattice model's regular part tends to the regular part of the zero-range model defined in Eqs.~(\ref{eq:danilov},\ref{eq:Phi_ZR}). Thus, in the zero-range limit:
\begin{multline}
\left( \frac{\partial E}{\partial(\ln R_t)}\right)_a=N(N-1)(N-2)
|\phi_0(\vn,\vn,\vn)|^2 \left( \frac{\partial h_0}{\partial(\ln R_t)}\right)_{\!b}
\\
\times \int d^3 r\, d^3 r_4\ldots d^3 r_N\,|B(\rr,\rr_4,\ldots,\rr_N)|^2.
\label{eq:proto_dEdRt}
\end{multline}
It remains to evaluate the derivative of $h_0$ with respect to $R_t$: This is achieved by applying (\ref{eq:proto_dEdRt}) to the case of an Efimov trimer in free space, where the regular part can be deduced from the known expression~\cite{CastinWernerTrimere} for the normalized wavefunction. This yields [Tab.~II, Eq.~(1)].

\subsection{Short-distance triplet distribution function}

Similarly to the pair distribution function $g^{(2)}$, one defines the triplet distribution function
$g^{(3)}(\rr_1,\rr_2,\rr_3) = \la \hat{\psi}^\dagger(\rr_1) \hat{\psi}^\dagger(\rr_2) 
\hat{\psi}^\dagger(\rr_3)
\hat{\psi}(\rr_3) \hat{\psi}(\rr_2) \hat{\psi}(\rr_1)\ra$, which is given in first quantization by $N (N-1) (N-2) \int d^3r_4\ldots d^3r_N\,|\psi(\rr_1,\ldots,\rr_N)|^2$.
In the limit $R\to0$ where the three positions $\rr_1,\rr_2,\rr_3$ approach each other, 
the many-body wavefunction behaves according to~(\ref{eq:danilov}).
The result~[Tab.~\ref{tab:bosons2}, Eq.~(2)], where the integral over $\cc_{123}$ is taken for fixed $\RR$
and $\Oom$,
then directly follows, using~[Tab.~\ref{tab:bosons2}, Eq.~(1)].
As a consequence, in a measurement of the positions of all the particles,
the mean number of triplets of particles having a small hyperradius $R$ is given by
\begin{multline}
N_{\rm triplets}(R<\epsilon) = \frac{1}{3!} \int_{R<\epsilon} d^3r_1 d^3 r_3 d^3 r_3 g^{(3)}(\rr_1,\rr_2,\rr_3) \\
\underset{\epsilon\to0}{\sim} \frac{m}{2\hbar^2|s_0|^2} \left(\frac{\partial E}{\partial (\ln R_t)}
\right)_a \epsilon^2 \left[1-\mathrm{Re}\, \frac{(\epsilon/R_t)^{2i|s_0|}}{1+i|s_0|}\right]
\label{eq:ntripl}
\end{multline}
where we used the Jacobian
$\frac{D(\rr_1,\rr_2,\rr_3)}{D(\cc_{123},\RR)}=3\sqrt{3}$ and the division by $3!$ takes into account the 
indistinguishability of the particles within a triplet.

\subsection{Decay rate due to three-body losses}

In experiments, the cold atomic gases are only
metastable: There exist deeply bound dimer states, that is with a binding energy 
of order $\hbar^2/(m b^2)$, where $b$ is the van der Waals length of the real atomic interaction.
These deeply bound states can be populated by three-body collisions, which are strongly exothermic (with
respect to the trapping potential depth) and thus lead to a net loss of atoms.
Usually, one expects that these deeply bound dimer states have a vanishing small effect on the metastable many-body states
for $b\to 0$; the metastable states then converge to stationary states 
described by the zero-range model.

In presence of the Efimov effect, however, the probability
$p_{\rm close}$ to find three particles within a distance $b$ (e.g., with an hyperradius $R<b$)
vanishes only as $b^2$ according to Eqs.~(\ref{eq:danilov},\ref{eq:Phi_ZR},\ref{eq:ntripl}). As the three-body loss rate
scales as $p_{\rm close} \hbar /m b^2$, it does not vanish in the zero-range limit \cite{PetrovPRL,EsryGammaEfi}. 
Fortunately, one can still in that limit simply include the losses by modifying the three-body boundary
conditions \cite{Braaten_etats_d_efimov,RevueBraaten2}: One keeps Eq.~(\ref{eq:danilov}) with a modified $\Phi$ deduced
from Eq.~(\ref{eq:Phi_ZR}) by the substitution
\begin{multline}
\sin\left[|s_0|\ln\frac{R}{R_t}\right] \rightarrow \frac{1}{2i} \left[e^{-\eta} e^{i|s_0|\ln(R/R_t)} \right. \\
\left. -e^{\eta}e^{-i|s_0|\ln(R/R_t)}\right].
\label{eq:cl_eta}
\end{multline}
The so-called inelasticity parameter $\eta \geq 0$ determines to which extent the reflection of the incoming hyperradial wave $\exp[-i|s_0|\ln(R/R_t)]$ 
on the point $R=0$ (where the 
non-universal short range three-body physics takes place) is elastic.

In this work, we have considered so far the ideal case where $\eta$ is strictly zero. We now show that this allows 
to access the decay rate due to three-body losses to first order in $\eta$ by taking simply a derivative of the loss-less
eigenenergies $E$. In a first approach, generalizing to three-body losses the procedure used for two-body losses
in \cite{Braaten}, we simply assume that $E(\ln R_t)$ 
is an analytic function of $\ln R_t$. As the substitution
(\ref{eq:cl_eta}) simply amounts to performing the change
\be
\ln R_t \rightarrow \ln R_t - \frac{i\eta}{|s_0|},
\ee
we conclude that the resulting eigenenergy for non-zero $\eta$ acquires an imaginary part
$-i \hbar \Gamma/2$ given to first order in $\eta$ by [Tab.~\ref{tab:bosons2}, Eq.~(3)].
Furthermore, we have developed an alternative approach, that relates for arbitrary $\eta$ the decay rate
$\Gamma$ to the integral of $|B|^2$, where $B$ is defined by Eq.~(\ref{eq:danilov}), see Appendix~\ref{app:Gamma}.
Combining this with [Tab.~\ref{tab:bosons2}, Eq.~(1)] in the limit $\eta\to 0$ 
reproduces the relation [Tab.~\ref{tab:bosons2}, Eq.~(3)].

{\bf Note added by F. Werner and Y. Castin after publication:}
As a simple application of relation [Tab. II, Eq.(3)] we calculate the decay rate $\Gamma$ of a free space 
Efimov trimer close to the atom-dimer dissociation threshold, to first order in the inelasticity 
parameter $\eta$. In the absence of three-body losses ($\eta=0$), the trimer internal 
energy $E$ behaves close to the threshold as (see Eq.~(51) of reference \cite{RevueBraaten2}):
\[
E \underset{a\to a_*^+}{=} -\frac{\hbar^2}{m a^2} \left[1 + D \ln^2\frac{a}{a_*}\right] +O(a-a_*)^3
\]
where $a_*$ is the value of the scattering length where the trimer energy meets the dimer energy, and 
the constant $D\simeq 0.164$ is related to one of the constants $c_i$ appearing in the expression of the atom-dimer 
scattering length $a_{\rm ad}=(c_1+c_2\cot [|s_0|\ln(a/a_*)])a$ by $D=3 |s_0|^2/(4 c_2^2)$. 
Since $a_*$ is proportional to the three-body parameter $R_t$ \cite{RevueBraaten2}, $da_*/d(\ln R_t)=a_*$ and
\[
\frac{d\Gamma}{d\eta}(\eta=0) \underset{a\to a_*^+}{\sim} \frac{4D}{|s_0|} \frac{\hbar}{ma_*^2} \ln \frac{a}{a_*}
\]
The decay rate (at small $\eta$) thus tends to zero, but more slowly than the binding energy $-\frac{\hbar^2}{ma^2}
-E$ of the trimer, leading to a vanishing quality factor at the dissociation threshold.

Our result disagrees with Eq.~(71) of reference \cite{RevueBraaten2}, or 
equivalently with Eq.~(259) of reference \cite{RevueBraaten},
where $\frac{d\Gamma}{d\eta}(\eta=0)$ has a nonzero limit
$\frac{8\hbar^2}{m a_*^2 |s_0|}$ at the dissociation threshold. This is due to the omission by the authors of
\cite{RevueBraaten,RevueBraaten2} 
of a term  proportional to $\Delta'(\xi)$, when they expand their Eq.~(69) in \cite{RevueBraaten2}
to first order in $\eta$ to obtain their Eq.~(71). Here $\Delta(\xi)$
is the so-called Efimov universal function, defined over the interval $[-\pi,-\pi/4]$, entering in the expression 
of the trimer energy, and $\Delta'(\xi)$ is its derivative. 
Performing the exact linearisation in $\eta$, we correct Eq.~(71) of reference \cite{RevueBraaten2} as:
\[
\frac{d\Gamma}{d\eta}(\eta=0)=\frac{4 |s_0|^{-1}\hbar^{-1}}{(\frac{\hbar^2}{ma^2}-E)^{-1} +
\frac{\Delta'(\xi)}{2|s_0|E} \frac{\tan \xi}{\tan' \xi}}
\]
with $\tan \xi=-a(-mE/\hbar^2)^{1/2}$. In the limit $a\to a_*^+$, this must reproduce our result, which
leads to the following property of the Efimov universal function:
\[
\Delta'(\xi) \underset{\xi\to (-\pi/4)^-}{\sim} \frac{2|s_0| D^{-1/2}}{|\xi+\frac{\pi}{4}|^{1/2}}
\]
{\bf End of the added note.}

\section{Application: Three-body loss rate for a Bose gas at thermal equilibrium}
\label{sec:tblr}

We consider a 3D Bose gas, in a cubic quantization box of volume $V$, at thermal equilibrium
in the grand canonical ensemble and in the thermodynamic limit. Within the zero-range model,
with a truncation of the three-body energy spectrum (that is introducing a lower energy cut-off,
as discussed below), relation [Tab.~\ref{tab:bosons2}, Eq.~(3)]
can be used to obtain, to first order in the inelasticity parameter $\eta$,
the three-body loss-constant $L_3$ customarily defined by
\be
\frac{d}{dt} N = - L_3 n^2 N
\ee
where $N$ is the mean particle number and $n=N/V$ the mean density.
Applying [Tab.~\ref{tab:bosons2}, Eq.~(3)] to each many-body eigenstate, taking a truncated
thermal average 
\footnote{To give a meaning to a $N$-body thermal average 
within the zero-range model requires, for $N\geq 4$, a procedure whose identification is beyond
the scope of this paper. This is here a formal issue, as we will consider the non-degenerate
limit allowing us to restrict to the three-body sector.}
and keeping in mind that each loss event eliminates three particles out of the system
\footnote{If one normalizes to unity the eigenstate $\psi$ at time $0$, 
the norm squared $||\psi(t)||^2$ is the probability that no loss event occured during $t$.
For the complex eigenenergy $E-i\hbar\Gamma/2$, this leads to a loss event rate equal to $\Gamma$,
and to a particle loss rate $dN/dt=-3\Gamma$.},
we obtain
\be
\frac{d L_3}{d\eta}(\eta=0) =  \frac{6}{\hbar |s_0| n^2 N} \left(\frac{\partial \Omega}{\partial(\ln R_t)}\right)_{\mu,T}
\ee
where the derivative of the grand potential $\Omega$ is taken for fixed chemical potential
$\mu$ and temperature $T$.

To obtain analytical results, we restrict to the non-degenerate
limit $\mu\to -\infty$, where the density vanishes, $n \lambda^3\to 0$, with $\lambda=[2\pi\hbar^2/(m k_B T)]^{1/2}$
the thermal de Broglie wavelength. One then can use the virial expansion 
\cite{Huang_livre,PaisUlhenbeck1959,BedaqueRupak2003,Ho_virial,Hu_virial3}:
\be
\Omega(\mu,T) = -\frac{V}{\lambda^3} k_B T \sum_{q\geq 1} b_q e^{q\beta\mu},
\ee
with $\beta=1/(k_B T)$, and $b_q$ only depends on $q$-body physics and temperature.
The leading order contribution that involves $\ln R_t$ is thus for $q=3$, so that
\be
\frac{d L_3}{d\eta}(\eta=0) \underset{n\lambda^3\to 0}{\to} -\frac{12\pi}{|s_0|} \frac{\hbar \lambda^4}{m} 
\left(\frac{\partial b_3}{\partial(\ln R_t)}\right)_{T}
\ee
where we used $n \lambda^3\sim \exp(\beta\mu)$.

The coefficient $b_q$ can be deduced from the solution of the $q$-body problem. We thus restrict
to the resonant case $1/a=0$, where the analytical solution for $q=3$ is known in free space \cite{Efimov}. 
Due to separability in hyperspherical coordinates \cite{WernerSym} the solution is also known
for the isotropic harmonic trap case \cite{Pethick3corps,Werner3corpsPRL}, which allows us to use the technique
developed in \cite{Hu_virial3,Huilong2010} to write $b_3$ as
\be
b_3 = 3^{3/2} \lim_{\omega\to 0} \left[\frac{Z_3}{Z_1} -Z_2 + \frac{1}{3} Z_1^2\right]
\ee
where $Z_q(\omega)$ is the canonical partition function at temperature $T$ for the system of $q$ interacting bosons
in the harmonic trapping potential $U(\rr)=\frac{1}{2} m\omega^2 r^2$.
Since the center-of-mass is separable, $Z_3/Z_1$ simply equals the partition function $Z_3^{\rm int}$
of the internal variables.
The internal 3-body eigenspectrum in the trap involves fully universal states (not depending on $R_t$),
and a single Efimovian channel with $R_t$-dependent eigenenergies $E_n(\omega)$, 
$n\in \mathbb{Z}$, solving a transcendental
equation.  Within the boundary conditions (\ref{eq:danilov},\ref{eq:Phi_ZR}), the sequence $E_n(\omega)$ 
is unbounded below. To give a mathematical existence
to thermal equilibrium, we thus truncate the sequence, labelling the ground three-body state
with the quantum number $n=0$ and then keeping only $n\geq 0$ in the thermal average
\footnote{Physically, our $n=0$ trimer state corresponds to 
the lowest weakly bound trimer. As usual in cold atom physics, the deeply bound (here trimer) states are 
excluded from the thermal ensemble since their (very exothermic) collisional formation 
simply leads to particle losses}. In the free space limit $\omega\to 0$, this corresponds to a purely geometric
spectrum of trimer states with a ratio $\exp(-2\pi/|s_0|)$ and a ground state Efimov trimer energy:
\be
E_0(\omega) \underset{\omega\to 0}{\to} -\frac{2\hbar^2}{m R_t^2} e^{\frac{2}{|s_0|} \mathrm{Im}\, \ln \Gamma (1+s_0)} \equiv - E_t.
\label{eq:defet}
\ee
Given $E_t$, this uniquely determines the three-body parameter $R_t$
\footnote{In reality, for an interaction with finite range or effective range $b$, 
the Efimovian trimer spectrum is only asymptotically geometric ($n\to +\infty$);
there exist various models \cite{stecher_Nbosons,Pricoupenko2010}, 
however, where $E_t$ is of order $\exp(-2\pi/|s_0|)\hbar^2/(m b^2)$ 
so that $R_t \gg b$, the ground state Efimovian trimer is close to the zero-range
limit, and the spectrum is almost entirely geometric.}.
This finally leads to
\be
\left(\frac{\partial b_3}{\partial(\ln R_t)}\right)_T = -\frac{3^{3/2}}{k_B T} \lim_{\omega\to 0} 
\sum_{n\geq 0} e^{-\beta E_n(\omega)}
\frac{\partial E_n(\omega)}{\partial(\ln R_t)}.
\label{eq:fsl}
\ee
Details of the calculation of that limit are exposed in Appendix~\ref{app:b3}. The resulting expression
for the three-body loss rate constant can be split in contributions of the three-body bound free-space spectrum
and continuous free-space spectrum:
\be
\frac{d L_3}{d\eta}(\eta=0) \underset{n\lambda^3\to 0}{\to}
72\sqrt{3}\, \frac{\hbar \lambda^4}{m} \left(S_{\rm bound} + S_{\rm cont}\right).
\label{eq:l3final}
\ee

The bound-state contribution naturally appears as a (rapidly converging) discrete sum over the trimer states:
\be
S_{\rm bound} = \frac{\pi}{|s_0|} \sum_{n\geq 0} \beta E_t  e^{-2\pi n/|s_0|} \exp\left(
\beta E_t  e^{-2\pi n/|s_0|}\right).
\label{eq:sbound}
\ee
This allows to predict the mean number $N_{\rm trim}$ of trimers with energy $E_{\rm trim}=-E_t  e^{-2\pi n/|s_0|}$ in the 
loss-less system at thermal equilibrium:
Since the contribution to $dN/dt$ (to first order in $\eta$) of the term of index $n$ in (\ref{eq:sbound}) is
intuitively $-3 \Gamma_{\rm trim} N_{\rm trim}$, where the decay rate of the trimer
is $\Gamma_{\rm trim}\simeq (2\eta/\hbar|s_0|) \partial_{\ln R_t} E_{\rm trim}$, we obtain
\be
\frac{N_{\rm trim}}{N} \underset{n\lambda^3\to 0}{\sim} 3^{3/2} (n \lambda^3)^2 e^{-\beta E_{\rm trim}}.
\ee
This agrees with Eq.~(188) of \cite{PaisUlhenbeck1959} obtained from a chemical equilibrium reasoning.

The continuous-spectrum contribution to (\ref{eq:l3final}) naturally appears as an integral over positive energies $E$,
see Appendix~\ref{app:b3}. Mathematically, it can also be turned into an easier to evaluate (rapidly
converging) discrete sum
\footnote{This is rapidly converging since $|\Gamma(1-in|s_0|)|^2=\pi n |s_0|/\sinh(\pi n |s_0|)$ \cite{Gradstein}.}:
\be
S_{\rm cont} = \frac{1}{2} + \sum_{n\geq 1} e^{-n \pi |s_0|} 
\mathrm{Re} \left[\Gamma(1-in|s_0|) \left(\beta E_t\right)^{i n |s_0|}\right].
\label{eq:scont}
\ee
As expected, $S_{\rm cont}$ is a log-periodic function of $E_t$. In practice, due to 
$|s_0|> 1$, it has weak amplitude
oscillations, between the extreme values $\simeq 0.478$ and $\simeq 0.522$.
Our continuous-spectrum contribution to $L_3$ is equivalent, to first order
in $\eta$,
to the result of a direct three-body loss rate calculation for the thermal ensemble of free-space three-boson
scattering states \cite{Petrov_manip_en_prepa}.

In experiments, the interaction potential has a finite range $b$, and the actual $L_3$ will 
deviate from the above results. For clarity, we now denote with a star
the quantities corresponding to a finite $b$. Due to the three-body losses, the so-called weakly bound
trimer states are actually not bound states, they are resonances with complex energies $E_n^*-i\hbar \Gamma_n^*/2$.
Assuming that $\Gamma_n^* \ll |E_n^*|$, we can name these resonances quasi-bound  states or
quasi-trimers.
Their contribution to the decay rate of the Bose gas,
from the reasoning below Eq.~(\ref{eq:sbound}), can be estimated as
\be
\Gamma_{\rm quasi-bound}^* \simeq
3^{3/2} (n \lambda^3)^2 N \sum_{n\geq 0} \Gamma_n^* e^{-\beta E_n^*}.
\ee
This is meaningful provided that the thermal equilibrium trimer
population formula Eq.~(188) of \cite{PaisUlhenbeck1959} makes sense in presence of losses,
that is the formation rate of quasi-trimers of quantum number $n$ has to remain much larger than $\Gamma_n^*$
(in the zero-range framework, this is ensured by first taking the limit $\eta\to 0$
and then the limit of vanishing density $n\lambda^3\to 0$).
Evaluation of the finite-$b$ positive-energy continuous spectrum 
contribution $L_{3, \mathrm{cont}>0}^*$  to the three-body loss rate constant is beyond the scope of this work. We can simply point out that, 
taking the limit $b\to 0$ (with a fixed, infinite scattering length) makes $L_{3, \mathrm{cont}>0}^*$ 
converge to the value obtained in the zero-range finite $\eta$ model; further taking the zero-$\eta$
limit gives
\be
\lim_{\eta\to 0} \frac{1}{\eta}\left(\lim_{b\to 0} L_{3, \mathrm{cont}>0}^*\right) = 
\frac{d L_{3,\mathrm{cont}}}{d\eta}(\eta=0).
\ee
In practice, as soon as $b\ll \lambda$ and $\eta\ll 1$, we expect that
$L_{3, \mathrm{cont}>0}^*\simeq \eta \frac{d L_{3,\mathrm{cont}}}{d\eta}(\eta=0)$.

\section{Arbitrary mixture}\label{sec:melange}

\begin{table*}
\begin{tabular}{|cc|cc|}
\hline  
Three dimensions &  & Two dimensions &  \\
\vspace{-4mm}
& & &  \\
\hline
\vspace{-4mm}
& & &  \\
$\displaystyle\frac{\partial E}{\partial(-1/a_{\si\sip})} = \frac{2\pi\hbar^2}{\mu_{\si\sip}} (A,A)_{\si\sip}$ &
(1a) &
$\ds\frac{\partial E}{\partial(\ln a_{\si\sip})} = \frac{\pi\hbar^2}{\mu_{\si\sip}} (A,A)_{\si\sip}$ &
(1b) \\
\vspace{-4mm}
& & & \\
\hline
\vspace{-4mm}
&&&\\
$\ds {C}_\si\equiv {\displaystyle \lim_{k\to +\infty}} 
k^4 n_\sigma(\kk) = 
\sum_{\sip}(1+\delta_{\si\sip})\frac{8\pi \mu_{\si\sip}}{\hbar^2}
\frac{\partial E}{\partial(-1/a_{\si \sip})} $ & (2a) &
$\ds {C}_\si\equiv {\displaystyle\lim_{k\to +\infty}} k^4 n_\sigma(\kk) = \sum_\sip(1+\delta_{\si\sip})\frac{4\pi \mu_{\si\sip}}{\hbar^2}
\frac{\partial E}{\partial(\ln a_{\si\sip})}$ & (2b) \\
\vspace{-4mm}
&&&\\
\hline
\vspace{-4mm}
& & &\\
$\ds
\int d^3c \,
 g_{\si\sip}^{(2)} \left(\mathbf{c}+\frac{m_\sip}{m_\si+m_\sip}\mathbf{r},
\mathbf{c}-\frac{m_\si}{m_\si+m_\sip}\mathbf{r}\right) $
&  &
$\ds
\int d^2c \,
 g_{\si\sip}^{(2)} \left(\mathbf{c}+\frac{m_\sip}{m_\si+m_\sip}\mathbf{r},
\mathbf{c}-\frac{m_\si}{m_\si+m_\sip}\mathbf{r}\right)
$
& 
\\ 
\vspace{-4mm}
& & &\\
 $\underset{r\to0}{\sim}(1+\delta_{\si\sip}) \ds \,\frac{\mu_{\si\sip}}{2\pi\hbar^2}\,\frac{\partial E}{\partial(-1/a_{\si\sip})}
\,\frac{1}{r^2}$ & (3a) &
$\underset{r\to0}{\sim}(1+\delta_{\si\sip}) \ds\, \frac{\mu_{\si\sip}}{\pi\hbar^2}\,\frac{\partial E}{\partial(\ln a_{\si\sip})}
\,\ln^2 r$ &(3b)
\\
\vspace{-4mm}
&&&\\
\hline
\vspace{-4mm}
& & & \\
$\ds E - E_{\rm trap}  = \sum_{\si\leq\sip} \frac{1}{a_{\si\sip}} \frac{\partial E}{\partial(-1/a_{\si\sip})}  $ &  &
$\ds E - E_{\rm trap}  = \lim_{\Lambda\to\infty}\left[-\sum_{\si\leq\sip}\frac{\partial E}{\partial(\ln a_{\si\sip})} \ln \left(\frac{a_{\si\sip} \Lambda e^\gamma}{2}\right) \right.
$ & 
\\
\vspace{-4mm}
& & & \\
 $\ds +\sum_{\sigma} \int \frac{d^3\!k}{(2\pi)^3}  \frac{\hbar^2 k^2}{2m_\si} 
\left[n_\sigma(\kk) - \frac{{C}_\si}{k^4}\right]$
& (4a) &
$\ds \left. +\sum_{\sigma} \int_{k<\Lambda} \frac{d^2\!k}{(2\pi)^2}  \frac{\hbar^2 k^2}{2m_\si} 
n_\sigma(\kk)\right]$ & (4b)
\\
\vspace{-4mm}
&&&\\
\hline
& & & \\
$\ds\frac{1}{2} \frac{\partial^2E_n}{\partial(-1/a_{\si\sip})^2}
= \left(\frac{2\pi\hbar^2}{\mu_{\si\sip}}\right)^2 \sum_{n',E_{n'}\neq E_n} 
\frac{|(A^{(n')},A^{(n)})_{\si\sip}|^2}{E_n-E_{n'}}$
& (5a) &
$\ds\frac{1}{2} \frac{\partial^2E_n}{\partial(\ln a_{\si\sip})^2}
= \left(\frac{\pi\hbar^2}{\mu_{\si\sip}}\right)^2 \sum_{n',E_{n'}\neq E_n} 
\frac{|(A^{(n')},A^{(n)})_{\si\sip}|^2}{E_n-E_{n'}} $
& (5b)
\\
\vspace{-4mm}
&& &\\
\hline 
\vspace{-4mm}
&& &\\
$\ds\left(\frac{\partial^2F}{\partial(-1/a_{\si\sip})^2}\right)_T < 0 $
& (6a) &
$\ds\left(\frac{\partial^2F}{\partial(\ln a_{\si\sip})^2}\right)_T < 0 $
& (6b)
\\
\vspace{-4mm}
& & & \\
\hline
\vspace{-4mm}
& & & \\
$\ds\left(\frac{\partial^2E}{\partial(-1/a_{\si\sip})^2}\right)_S < 0 $
& (7a) &
$\ds\left(\frac{\partial^2E}{\partial(\ln a_{\si\sip})^2}\right)_S < 0 $
& (7b) \vspace{-4mm}
\\
& & & \\
\hline
\end{tabular}
\caption{Main results for an arbitrary mixture. 
In three dimensions, if the Efimov effect occurs, the derivatives must be taken for fixed three-body parameter(s), the expression for $E$ in line~4 breaks down,
and the last two lines, with derivatives of the free energy $F$ and of the mean energy $E$ 
respectively taken at fixed temperature $T$ and entropy $S$,
are meaningless in the absence of spectral selection (see
Sec.~\ref{sec:tblr}). $\gamma=0.577215\ldots$ is Euler's constant.
\label{tab:melange}}
\end{table*}


In this Section we consider a mixture of bosonic and/or fermionic atoms with an arbitrary number of spin components. The $N$ particles are thus divided into groups, each group corresponding to a given chemical species and to a given spin state. We label these groups by an integer $\sigma\in\{1,\ldots,n\}$.
Assuming that there are no spin-changing collisions, the number $N_\sigma$ of atoms in each group is fixed, and one can consider that particle $i$ belongs to the group $\sigma$ if $i\in I_\sigma$, where the $I_\sigma$'s are a fixed partition of
$\{1,\ldots,N\}$ which can be chosen arbitrarily. For example, 
a possible choice is $I_1=\{1,\ldots,N_1\}$; $I_2=\{N_1+1,\ldots,N_1+N_2\}$; etc.
 The wavefunction $\psi(\rr_1,\ldots,\rr_N)$ is then symmetric (resp. antisymmetric) with respect to the exchange of two particles  belonging to the same group $I_\sigma$ of bosonic (resp. fermionic) particles.
Each atom has a mass $m_i$ and is subject to a trapping potential $U_i(\rr_i)$, and the scattering length between atoms $i$ and $j$ is $a_{ij}$.  We set $m_i=m_\sigma$ and $a_{ij}=a_{\sigma \sigma'}$
for $i\in I_\sigma$ and $j\in I_{\sigma'}$. The reduced masses are $\mu_{\si\sip}=m_\si m_\sip/(m_\si+m_\sip)$.
We shall denote by $\mathcal{P}_{\sigma \sigma'}$ the set of all pairs of particles with one particle in group $\sigma$ and the other one in group $\sigma'$, each pair being counted only once:
\be
\mathcal{P}_{\sigma \sigma'}\equiv\left\{ (i,j)\in (I_\sigma\times I_{\sigma'})\cup (I_{\sigma'}\times I_\sigma) \ /\ i<j \right\}.
\ee

The definition of the zero-range model is modified as follows: In the contact conditions~(\ref{eq:wbp3d},\ref{eq:wbp2d}),
the scattering length $a$ is replaced by $a_{ij}$,
and the limit $r_{ij}\to0$ is taken for a fixed center of mass position $\cc_{ij}=(m_i\rr_i+m_j\rr_j)/(m_i+m_j)$;
moreover Schr\"odinger's equation becomes
\be
\sum_{i=1}^N \left[
-\frac{\hbar^2}{2 m_i}\Delta_{\rr_i} + U_i(\rr_i)
\right] \psi = E\,\psi.
\ee

Our results are summarized in Table~\ref{tab:melange}, where we introduced the notation in dimension $d$:
\begin{multline}
( A^{(1)},A^{(2)})_{\si\sip}\equiv \sum_{(i,j)\in \mathcal{P}_{\si \sip}} \int \Big( \prod_{k\neq i,j} d^d r_k \Big) \int d^d c_{ij}
\\
A^{(1)*}_{ij}(\mathbf{c}_{ij}, (\mathbf{r}_k)_{k\neq i,j})
A^{(2)}_{ij}(\mathbf{c}_{ij}, (\mathbf{r}_k)_{k\neq i,j}).
\end{multline}
Since $a_{\si\sip}=a_{\sip\si}$ there are only $n(n+1)/2$ independent scattering lengths, and the 
 partial derivatives with respect to one of these independent scattering lengths are taken while keeping fixed the other independent scattering lengths.
We note that, in Ref.~\cite{CombescotC}, [Tab.~\ref{tab:melange}, Eqs.~(4a,4b)]
were already partially obtained
\footnote{Our expressions [Tab.~\ref{tab:melange}, Eqs.~(4a,4b)] complete the ones in~\cite{CombescotC} in the following way. 
In Ref.~\cite{CombescotC},the coefficient of $1/a_{\sigma\sigma'}$ 
was not expressed as $\partial E/\partial(1/a_{\sigma\sigma'})$; 
only the case of a spatially homogeneous system 
was covered; finally, an arbitrary mixture was covered only in 3D, while in 2D only the 
case of a 2-component Fermi-Fermi mixture was covered.}.

In 3D the three-body Efimov effect occurs, except for a mixture of only two fermionic groups with a heavy-to-light mass ratio 
$m_\sigma/m_{\sigma'}<13.6069\ldots$ \cite{Efimov73,Petrov3fermions,LudoYvanBoite}. When the three-body Efimov effect occurs, as for single-component bosons, the derivatives with respect to any scattering length have {\sl a minima} to be taken for fixed three-body parameter(s), and the relation between $E$ and the momentum distribution [Tab.~\ref{tab:melange}, Eq.~(4a)] breaks down, which was not realized in~\cite{CombescotC}~\footnote{Indeed, in presence of the Efimov effect, the momentum distribution has a subleading contribution 
$\delta n_\si(k)$ scaling as $1/k^5$, evaluated in the bosonic case in~\cite{CastinWerner_nk_trimer}, leading to a divergent integral in this relation. For two-component fermions with a mass ratio sufficiently close to $1$,
the integral converges, because $\delta n_\si(k)\propto 1/k^{5+2s}$ where $s>0$ is the scaling exponent of the three-body wavefunction, $\psi(\lambda\rr_1,\lambda\rr_2,\lambda\rr_3)\propto\lambda^{s-2}$ for $\lambda\to0$, see a note in~\cite{TanLargeMomentum} and note 6 in \cite{CompanionFermions}.};
moreover, we expect new relations analogous to the ones given in Section~\ref{sec:new_rel} for bosons.
Furthermore, we assume here that there is no fermionic group $\sigma$ with a mass ratio $m_\sigma/m_{\sigma'}>13.384$ with respect to any other group
$\sigma'$, so as to avoid a four-body Efimov effect \cite{CMP}. More generally, the zero-range model Hamiltonian is assumed to be self-adjoint
without introducing interaction parameters other than scattering lengths and three-body parameters.


The derivations of the relations of Tab.~\ref{tab:melange} are analogous to the ones already given for two-component fermions and single-component bosons. 
The lemmas~[Article~I, Eqs.~(33,35)] are replaced by
\begin{multline}
\la\psi_1,H\psi_2\ra-\la H\psi_1,\psi_2\ra
\\
=
\left\{
\begin{array}{lr}
\ds \frac{2\pi\hbar^2}{\mu_{\si\sip}}\left(\frac{1}{a^{(1)}_{\sigma\sigma'}}-\frac{1}{a^{(2)}_{\sigma\sigma'}}\right)(A^{(1)},A^{(2)})_{\si\sip} & {\rm in}\ 3D
\\
\ds\frac{\pi\hbar^2}{\mu_{\si\sip}}\ln(a^{(2)}_{\sigma\sigma'}/a^{(1)}_{\sigma\sigma'})(A^{(1)},A^{(2)})_{\si\sip} & {\rm in}\ 2D,
\end{array}
\right.
\end{multline}
where $\psi_1$ and $\psi_2$ obey the same contact conditions (including the three-body ones
if there is an Efimov effect), {\sl except} for the independent scattering length $a_{\sigma\sigma'}$,
that is equal to $a^{(i)}_{\sigma\sigma'}$ for $\psi_i$, $i=1,2$.
The momentum distribution for the goup $\sigma$ is normalized as
$\int n_\sigma(\kk) d^dk/(2\pi)^d=N_\sigma$.
The pair distribution function is now defined by
\begin{multline}
g^{(2)}_{\si\sip}(\mathbf{u},\mathbf{v})
=
\int d^dr_1\ldots d^dr_N 
\left|
\psi(\rr_1,\ldots,\rr_N)
\right|^2
\\
\times \sum_{ i\in I_\si , j\in I_\sip , i\neq j}
\delta\left(\mathbf{u}-\rr_i\right)
\delta\left(\mathbf{v}-\rr_j\right).
\end{multline}
The Hamiltonian of the lattice model used in some of the derivations now reads
\be
H_{\rm latt}=H_0+\sum_{\sigma\leq\sigma'}g_{0,\sigma \sigma'} \,W_{\sigma \sigma'}
\ee
where
$
H_0=\sum_{i=1}^N \left[ -\frac{\hbar^2}{2m_i}\Delta_{\rr_i} +U_i(\rr_i) \right]
$
with the discrete Laplacian defined by $\langle \rr | \Delta_\rr | \kk \rangle \equiv -k^2 \langle \rr | \kk \rangle$ (for $\kk$ in the first Brillouin zone)
and
$W_{\sigma \sigma'}=\sum_{(i,j)\in \mathcal{P}_{\sigma \sigma'}} \delta_{\rr_i,\rr_j} b^{-d}.$
In the formulas of Article~I
involving the two-body scattering problem, one has to replace $g_0$ by $g_{0,\sigma\sigma'}$, $a$ by $a_{\si\sip}$ and $m$ by $2\mu_{\si\sip}$. Denoting the corresponding zero-energy scattering wavefunction by $\phi_{\si\si'}(\rr)$,
the lemma~[Article~I, Eq.~(56)] is replaced by
$\la\psi'|W_{\si\sip}|\psi\ra = |\phi_{\si\sip}({\bf 0})|^2\ ( A',A)_{\si\sip}.$

\section{Conclusion} \label{sec:conclusion}

In dimensions two and three,
we obtained several relations valid for any eigenstate of the $N$-boson problem with zero-range interactions.
The interactions are characterized by the 2D or 3D  two-body $s$-wave 
scattering length $a$ and, in 3D when the Efimov effect 
takes place, 
by a three-body parameter $R_t$. Our expressions relate various observables to derivatives of the energy
with respect to these interaction parameters. Some of the expressions, 
initially obtained in \cite{50pages},
were derived in \cite{BraatenBosons} with a different technique. For completeness, we
have also generalized some of the relations to arbitrary mixtures of Bose and/or Fermi gases.

For the bosons in 3D, especially interesting are the relations involving the 
derivative of the energy with respect to the three-body parameter.
Physically, one of then predicts (to first order in the inelasticity parameter $\eta$)
the decay rate $\Gamma$ of the system due to three-body losses, that occur
in cold atom experiments by recombination to deeply bound dimers. This means that
one can extract $\Gamma$ from the eigenenergies of a purely loss-less ($\eta=0$) model.  As an application,
we analytically obtained (within the zero-range model, and to first order in $\eta$) 
the three-body loss rate constant $L_3$ for the 3D non-degenerate Bose gas at thermal equilibrium with infinite
scattering length. Experimentally, this quantity is under current study with real atomic gases 
\cite{Petrov_manip_en_prepa}.

Mathematically, the 3D relations hold under the assumption that the two-body scattering length and the three-body parameter are sufficient to make the $N$-boson problem well-defined,  with a self-adjoint Hamiltonian.
Therefore they may be used to numerically test this assumption, for example by checking the consistency between
the values of
the derivative of the energy with respect to the three-body parameter
obtained in different ways. Three possible ways are: numerical differentiation of the energy,
the present relation on the short-distance triplet distribution function,
or the virial theorem which also involves this derivative~\cite{FelixViriel}.

\acknowledgments
We thank S.~Tan and J.~von~Stecher  for stimulating discussions.
The work of F.W. at UMass was supported by NSF under Grant No.~PHY-0653183 and No.~PHY-1005543.
Our group at ENS is a member of IFRAF. We acknowledge support from ERC
Project FERLODIM N.228177.

\appendix

\section{Derivation of a lemma}
\label{app:3b}

Here we derive the lemma (\ref{eq:lemme_dEdRt}) for three bosons in the zero-range model.
The first step is to express the Hamiltonian in hyperspherical coordinates~\cite{WernerThese,LeChapitre}:
Using the value of the Jacobian given below Eq.~(\ref{eq:ntripl}),
\begin{multline}
\la \psi_1, H \psi_2\ra-\la H\psi_1,\psi_2\ra
\\ =
-\frac{\hbar^2}{2m}3\sqrt{3} \int_0^\infty dR\,R^5 \int d^5\Omega \int d^3c
\\ 
 \left\{\psi^*_1\left(\frac{\partial^2}{\partial R^2}+\frac{5}{R}\frac{\partial}{\partial R}
+\frac{T_\Oom}{R^2} +\frac{1}{3}\Delta_\cc\right)\psi_2
-\left[\psi^*_1 \leftrightarrow \psi_2
\right]
\right\} 
\\
=
-\frac{\hbar^2}{2m}3\sqrt{3}
\left\{\int_0^\infty dR\,R^5 \int d^5\Omega\,\cA_\cc(R,\Oom)
\right.
\\
+\int d^5\Omega\, d^3c\,\cA_R(\Oom,\cc)
\left.
+\int_0^\infty dR\,R^5\int d^3c\,\cA_\Oom(R,\cc)\right\}
\end{multline}
where $\cc=\cc_{123}$ and
\bea
\cA_\cc(R,\Oom)&\equiv&\int d^3c\,\left\{\psi^*_1\,\frac{1}{3}\Delta_\cc\,\psi_2
-\left[\psi^*_1 \leftrightarrow \psi_2
\right]
\right\}
\\
\cA_R(\Oom,\cc)&\equiv&\int_0^\infty dR\,R^5\Bigg\{\psi^*_1\left(\frac{\partial^2}{\partial R^2}+\frac{5}{R}\frac{\partial}{\partial R}\right)\psi_2
\nonumber
\\ & &
-\left[\psi^*_1 \leftrightarrow \psi_2
\right]\Bigg\}
\\
\cA_\Oom(R,\cc)&\equiv&\int d^5\Omega\left\{\psi^*_1 \frac{T_\Oom}{R^2}\psi_2
-\psi_2 \frac{T_\Oom}{R^2}\psi^*_1\right\},
\eea
$T_\Oom$ being a differential operator acting on the hyperangles and called Laplacian on the hypersphere.

The quantity $\cA_R$ can be computed using the following simple lemma: If $\Phi_1(R)$ and $\Phi_2(R)$ are functions which decay quickly at infinity and have no singularity except maybe at $R=0$, then
\begin{multline}
\int_0^\infty dR\,R^5\left\{\Phi_1^*\left(\frac{\partial^2}{\partial R^2}
+\frac{5}{R}\frac{\partial}{\partial R}\right)\Phi_2
-\left[\Phi_1^* \leftrightarrow \Phi_2\right]\right\}
\\
=
-\lim_{R\to0} R\left( \mathcal{F}_1^* \frac{\partial \mathcal{F}_2}{\partial R} - \mathcal{F}_2 \frac{\partial \mathcal{F}_1^*}{\partial R} \right)
\label{eq:lemmeR}
\end{multline}
where $\mathcal{F}_i(R)\equiv R^2\,\Phi_i(R)$. 
Expressing the right-hand-side of (\ref{eq:lemmeR}) thanks to the boundary condition (\ref{eq:danilov}) then yields the desired result (\ref{eq:lemme_dEdRt}), because the other two contributions $\cA_{\cc}$ and $\cA_\Oom$ both vanish as we now show.

The quantity $\cA_\cc(R,\Oom)$, rewritten as $\frac{1}{3}\int d^3c\, \nabla_\cc \cdot \left( \psi^*_1 \nabla_\cc \psi_2
- \psi_2 \nabla_\cc \psi^*_1 \right)$ with the divergence theorem, is zero, since the $\psi_i$'s are regular functions of $\cc$ for every $(R,\Oom)$ except on a set of measure zero.

It remains to show that
\be
\cA_\Oom(R,\cc)=0\  {\rm for\ any}\ \cc\ {\rm and\ } R>0.
\label{eq:A_Om=0}
\ee
We will use the fact that $\psi_1$ and $\psi_2$ satisfy the two-body boundary condition 
(\ref{eq:wbp3d}) with the same $a$, and apply lemma [Article~I, Eq.~(33)]. More precisely, we will show that for any smooth function $f(R,\cc)$ which vanishes in a neighborhood of $R=0$,
\be
\int_0^\infty dR\,R^5 \int d^3c \, f(R,\cc)^2\,\cA_\Oom(R,\cc)=0;
\label{eq:lemme_f}
\ee
this clearly implies (\ref{eq:A_Om=0}).
To show (\ref{eq:lemme_f}) we note that
\begin{multline}
-\frac{\hbar^2}{2m}3\sqrt{3}
\int_0^\infty dR\,R^5 \int d^3c \, f(R,\cc)^2\,\cA_\Oom(R,\cc)
\\=
-\frac{\hbar^2}{2m}3\sqrt{3}
\int_0^\infty dR\,R^5\int d^5\Omega\int d^3c
\\
\left\{ (f \psi_1^*)\frac{T_\Oom}{R^2} (f \psi_2)
-\left[\psi^*_1 \leftrightarrow \psi_2\right] \right\},
\end{multline}
which can be rewritten as
\begin{multline}
\int d^3r_1 d^3r_2 d^3r_3
\left\{ (f \psi_1^*)H (f \psi_2)
-\left[\psi^*_1 \leftrightarrow \psi_2 \right] \right\}
\\
+\frac{\hbar^2}{2m}3\sqrt{3}
\int_0^\infty dR\,R^5 \int d^5\Omega \int d^3c
\\ 
\left\{ (f \psi_1^*)
\left(\frac{\partial^2}{\partial R^2}+\frac{5}{R}\frac{\partial}{\partial R}
 +\frac{1}{3}\Delta_\cc\right) (f \psi_2)
-\left[\psi^*_1 \leftrightarrow \psi_2 \right]\right\}.
\label{eq:expression}
\end{multline}
The first integral in this expression vanishes, as a consequence of the lemma [Article~I, Eq.~(33)].
This lemma is indeed applicable to the wavefunctions $f\psi_i$:
They vanish in a neighborhood of $R=0$ (see the discussion in Article~I), moreover
 they satisfy the two-body boundary condition for the same value of the scattering length $a$ (as follows from the fact that $R$ varies quadratically with $r$ for small $r$).
 The second integral in (\ref{eq:expression}) vanishes as well: The contribution of the partial derivatives with respect to $R$ vanishes as a consequence of lemma (\ref{eq:lemmeR}), and the contribution of $\Delta_\cc$ vanishes because the $f\psi_i$'s are regular functions of $\cc$.

\section{Relation between $\Gamma$ and $B$ for any $\eta$}
\label{app:Gamma}

Contrarily to the remaining part of the paper, we assume here that the inelasticity
parameter $\eta>0$ and is not necessarily a small perturbation, so that the $N$-body wavefunction $\psi$ obeys contact conditions
given by Eq.(\ref{eq:danilov}) and by Eq.(\ref{eq:Phi_ZR}) {\sl modified} according to
(\ref{eq:cl_eta}). As a consequence, $\psi$ is in general an eigenstate of $H$ with a complex
energy $E-i\hbar \Gamma/2$, where $\Gamma$ is the decay rate. If $\psi$ is normalized to unity at time $0$ then
\be
\Gamma  = -\frac{d}{dt} (||\psi||^2)(t=0).
\ee
This can be transformed using the continuity equation
\be
\partial_t |\psi(\XX,t)|^2 + \mathrm{div}_{\XX} \JJ = 0
\label{eq:cont}
\ee
where we collected all the particles coordinates in a single vector
$\XX=(\rr_1,\ldots,\rr_N)$ with $3N$ components, and where we introduced
the probability current in $\mathbb{R}^{3N}$:
\be
\JJ = \frac{\hbar}{m} \mathrm{Im}\, (\psi^* \mathrm{grad}_\XX \psi).
\ee
Eq.~(\ref{eq:cont}) is valid for all $r_{ij}>0$, and results as usual from Schr\"odinger's equation.

To avoid the singularities that appear in $\psi$ for three coinciding particle positions,
we introduce exclusion volumes
$B_{ijk}(\epsilon) = \{ \XX\in\mathbb{R}^{3N} / R_{ijk} < \epsilon\}$
for all triplets $\{i,j,k\}$ of particles (of hyperradius $R_{ijk}$)
in the integral defining $||\psi||^2$, taking the limit $\epsilon\to 0$ at the end
of the calculation. With the divergence theorem, this leads to
\bea
\Gamma &=& -\lim_{\epsilon\to 0} \int_{I_\epsilon} d^{3N}X\, \partial_t (|\psi(\XX,t=0)|^2)  \nonumber\\
&=& -\lim_{\epsilon\to 0} \sum_{\{i,j,k\}} \int_{\partial B_{ijk}(\epsilon)} d^{3N-1}\SSS\cdot\JJ
\eea
with the surface element $d^{3N-1}\SSS$ oriented towards the exterior of $B_{ijk}$. 
Here $I_\epsilon$ is $\mathbb{R}^{3N}$ minus the union of all $B_{ijk}(\epsilon)$;
it is thus the set of all the $\XX$ having all the $R_{ijk}>\epsilon$.
Using the bosonic symmetry we single out the decay rate due to particles 1, 2 and 3:
\be
\Gamma = -\frac{N(N-1)(N-2)}{3!} \lim_{\epsilon\to 0}  \int_{\partial B_{123}(\epsilon)} d^{3N-1}\SSS\cdot \JJ.
\label{eq:cdd}
\ee
The integration domain in Eq.~(\ref{eq:cdd}), which is the boundary of $B_{123}(\epsilon)$,
is actually a cylinder in $\mathbb{R}^{3N}$, and
the coordinates number $10$ to $3N$ of the surface element $d^{3N-1}\SSS$ are zero, so that one
can keep the contribution to the probability current of the first 3 particles only:
We can thus replace $d^{3N-1}\SSS\cdot\JJ$ with $d^8\SSS_t\cdot\JJ_t$, the nine-coordinate
vectors $\JJ_t$ and $d^8\SSS_t$ coinciding with the first nine coordinates of $\JJ$ 
and $d^{3N-1}\SSS$.  For fixed $\rr_4,\ldots,\rr_N$ we thus have to evaluate
\be
\gamma(\epsilon) \equiv -\int_{R=\epsilon} d^8\SSS_t\cdot \JJ_t=\int_{R>\epsilon} 
d^3r_1 d^3r_2 d^3r_3 \mathrm{div}_{\rr_1,\rr_2,\rr_3} \JJ_t,
\ee
where we used the divergence theorem.
We then change the integration variables from $\rr_1,\rr_2,\rr_3$ to $\cc_{123},\RR$,
with a Jacobian given below Eq.~(\ref{eq:ntripl}).
Further use of the identity
\begin{multline}
\sum_{i=1}^{3} \mathrm{div}_{\rr_i} \left(\psi^*\mathrm{grad}_{\rr_i}\psi-\mbox{c.c.}\right)
=
\mathrm{div}_{\RR} \left(\psi^*\mathrm{grad}_{\RR}\psi-\mbox{c.c.}\right) \\
+
\frac{1}{3} \mathrm{div}_{\cc_{123}} \left(\psi^*\mathrm{grad}_{\cc_{123}}\psi-\mbox{c.c.}\right)
\end{multline}
and backward application of the divergence theorem yields
\be
\gamma(\epsilon)=-3\sqrt{3}\,\epsilon^5  \int d^3c_{123} \int d^5\Omega\, \frac{\hbar}{m} \mathrm{Im}
[\psi^*\partial_R \psi]_{R=\epsilon}.
\ee
The $R\to 0$ behavior of $\psi$ being given by $B$ times a known function, see Eq.~(\ref{eq:danilov}) and Eq.(\ref{eq:Phi_ZR}) 
modified according to (\ref{eq:cl_eta}), we finally obtain
\be
\Gamma = \frac{\hbar}{m} N(N-1)(N-2) \frac{\sqrt{3}}{4} |s_0| \sinh(2\eta) ||B||^2
\ee
with $||B||^2=\int d^3c_{123}\, d^3r_4\ldots d^3r_N\,|B(\cc_{123},\rr_4,\ldots,\rr_N)|^2$.
In the limit $\eta\to 0$, $||B||^2$ tends to its value in the loss-less model
and we recover [Tab.~\ref{tab:bosons2}, Eq.~(3)] using [Tab.~\ref{tab:bosons2}, Eq.~(1)].

\section{Free space limit of a virial sum}
\label{app:b3}

Here we derive the free-space limit (\ref{eq:fsl}) of a sum over the internal Efimovian
eigenenergies $E_n(\omega)$ for three bosons in a harmonic trap with oscillation frequency $\omega$,
interacting in the zero-range limit with infinite scattering length.
A rewriting of the implicit equation for $E_n$ of \cite{Pethick3corps} gives,
for $n\in\mathbb{N}$:
\be
\label{eq:impli}
\mathrm{Im} \ln \Gamma\left(\frac{1+s_0-\tilde{E}_n}{2}\right)
+\frac{|s_0|}{2} \ln \left(\frac{2\hbar\omega}{E_t}\right) + n \pi =0.
\ee
We have introduced the notation $\tilde{E}_n=E_n/(\hbar\omega)$.
Also, $\Gamma(z)$ is the Gamma function and its logarithm $\ln \Gamma(z)$ is the usual univalued function
with a branch cut on the real negative axis. The left-hand side of (\ref{eq:impli}) can be shown 
to be a decreasing function of $E_n$, using relation 8.362(1) of \cite{Gradstein}, 
so that Eq.~(\ref{eq:impli}) determines $E_n$ in a unique way.
The fact that $E_t$, as given by (\ref{eq:defet}), is the free space ground trimer binding energy
can be checked from (\ref{eq:impli}) by a Stirling expansion for $\tilde{E}_n\to -\infty$.

To evaluate the sum in (\ref{eq:fsl}) for $\omega\to 0$, we collect the eigenenergies $E_n$ into
three groups. The (finite) {\sl transition} group corresponds to $|E_n|$ not much larger than 
$\hbar \omega$, and
gives a vanishing contribution to (\ref{eq:fsl}) for $\omega\to 0$.
The {\sl bound state} group corresponds to negative eigenenergies with $|E_n|\gg \hbar \omega$; the corresponding free space
trimer sizes are much smaller than the harmonic oscillator length $[\hbar/(m\omega)]^{1/2}$, so that the trapping potential 
has a negligible effect and $E_n(\omega)$ is close to the free space trimer energy of quantum number $n$:
\be
E_n(\omega) \simeq -E_t e^{-2\pi n/|s_0|}.
\ee
This directly leads to the contribution $S_{\rm bound}$ in (\ref{eq:sbound}).

Finally, the third group contains the positive eigenenergies with $E_n \gg \hbar \omega$, that shall
reconstruct the free space continuous spectrum for $\omega\to 0$. As shown in Sec.~3.3.a of \cite{WernerThese},
these $E_n$ are almost equally spaced by $2\hbar \omega$. We need here the leading order deviation
from equispacing, that we parametrize with a ``quantum defect" $\Delta$ as
\be
\tilde{E}_n \underset{n\to +\infty}{=} 2n + \Delta(E_n) + O(1/n).
\ee
For $\tilde{E}_n\to +\infty$, Stirling's formula cannot be immediately applied to (\ref{eq:impli})
since the argument of the Gamma function remains at finite distance
from the poles of $\Gamma$ (on the real negative axis). Using $\Gamma(z)\Gamma(1-z)=\pi/\sin(\pi z)$ \cite{Gradstein},
we obtain the useful identity:
\begin{multline}
-\mathrm{Im}\, \ln \Gamma\left(\frac{1+s_0-\tilde{E}}{2}\right)=
\mathrm{Im}\, \ln \Gamma\left(\frac{1-s_0+\tilde{E}}{2}\right) \\
+\frac{\pi}{2} \tilde{E} +\mathrm{Im}\,\ln \left[1+e^{-\pi|s_0|}e^{-i\pi\tilde{E}}\right]
\label{eq:usid}
\end{multline}
for all real $\tilde{E}$. Note that the logarithm in the last term of that expression is unambiguously
defined (by a series expansion of $\ln(1+u)$ with $u$) since $e^{-\pi |s_0|}<1$. Stirling's expansion
can now be used in the right-hand side of (\ref{eq:usid}), turning (\ref{eq:impli}) into
an implicit equation for the ``quantum defect" $\Delta$:
\be
\Delta(E) = \frac{|s_0|}{\pi} \ln\left(\frac{E}{E_t}\right) -\frac{2}{\pi}
\mathrm{Im}\, \ln \left[1+e^{-\pi|s_0|} e^{-i\pi\Delta(E)}\right].
\label{eq:implidelta}
\ee
Since $\exp(-\pi|s_0|)\ll 1$, we have a {\sl small-deviation property}:
$\Delta(E)$ only slightly deviates, by $O[\exp(-\pi|s_0|)]$, from
the first term in the right-hand side of (\ref{eq:implidelta}). This deviation was not fully taken into
account in \S3.3.a of \cite{WernerThese}.  To remain exact, we multiply (\ref{eq:implidelta}) by $i\pi$
on both sides, and we exponentiate the resulting equation. Since $\exp[-2i\mathrm{Im}\, \ln(1+u)]=
(1+u^*)/(1+u)$, we obtain a solvable equation for $\exp(i\pi\Delta)$ that determines
$\Delta$ modulo $2$.  From the {\sl small-deviation
property} stated above, we can lift the modulo $2$ uncertainty:
\be
\Delta(E) = \frac{|s_0|}{\pi} \ln\left(\frac{E}{E_t}\right) +\frac{2}{\pi}
\mathrm{Im}\, \ln \left[1-e^{-\pi|s_0|} \left(\frac{E}{E_t}\right)^{-i|s_0|}\right].
\ee
Finally, it remains in (\ref{eq:fsl}) to replace the sum over $n$ (for $E_n$ in the third group)
by an integral $\int_0^{+\infty} dE/(2\hbar \omega)$, where $2\hbar\omega$ is the leading order
level spacing, to obtain the continuous spectrum contribution
\be
\left(\frac{\partial b_3}{\partial(\ln R_t)}\right)_T^{\rm cont} = -\frac{3^{3/2}}{2 k_B T}
\int_0^{+\infty} dE\, e^{-\beta E} \frac{\partial \Delta(E)}{\partial(\ln R_t)}.
\ee
After expansion of $\partial_{\ln R_t} \Delta(E)$ in powers of $e^{-\pi |s_0|}$, the integral over $E$ 
can be expressed in terms of the Gamma function, which eventually leads
to (\ref{eq:scont}).


\end{document}